\documentclass{article}
\usepackage{spconf}
\usepackage{amsmath,graphicx}
\usepackage{amssymb,amsfonts}
\usepackage{bm}
\usepackage{cite}
\usepackage{algorithm}
\usepackage{algorithmic}
\usepackage{enumerate}
\usepackage{url}
\usepackage{hyperref}
\usepackage{subfigure}

\makeatletter
\renewcommand\section{\@startsection{section}{1}{\z@}%
  {-3.0ex plus -1ex minus -.2ex}%
  {0.6ex plus .1ex}%
  {\normalfont\large\bfseries}}
\renewcommand\subsection{\@startsection{subsection}{2}{\z@}%
  {-2.0ex plus -0.8ex minus -.2ex}%
  {0.4ex plus .1ex}%
  {\normalfont\normalsize\bfseries}}
\def\name#1{\gdef\@name{{\em #1}}}
\def\@maketitle{\newpage
 \null
 \vskip 2em \begin{center}
 {\large \bf \@title \par} \vskip 1.5em {\large \lineskip .5em
\begin{tabular}[t]{c}\@name \\[0.5em] \@address
 \end{tabular}\par} \end{center}
 \par
 \vskip 0.3em}
\makeatother

\DeclareMathOperator{\diag}{diag}
\DeclareMathOperator*{\argmin}{arg\,min}

\newtheorem{assumption}{Assumption}

\newtheorem{theorem}{Theorem}

\title{Robust High-Dimensional MVDR Beamforming under Heavy-Tailed Noise via Spiked Covariance Modeling}
\name{Liusha Yang\textsuperscript{1}, Shuqi Chai\textsuperscript{2*}, Manyou Ma\textsuperscript{1*}\thanks{*Corresponding author}}
\address{\textsuperscript{1}School of Artificial Intelligence, Shenzhen Technology University \\\textsuperscript{2}Shenzhen Research Institute of Big Data}
\begin{document}
\topmargin=0mm 
\ninept
\maketitle

\begin{abstract}
\vspace{-0.1cm}
This paper proposes a robust high-dimensional minimum variance distortionless response (MVDR) beamforming method for array observations corrupted by heavy-tailed noise. The proposed approach constructs an MVDR-oriented precision matrix estimator by combining Maronna's robust scatter estimator with spiked covariance modeling. Using tools from random matrix theory, we derive a deterministic equivalent of the MVDR output power and obtain asymptotically optimal shrinkage eigenvalues for the dominant signal subspace. A fully sample-based implementation is then developed for practical beamformer design. Numerical simulations under elliptically distributed noise demonstrate that the proposed beamformer achieves stronger interference suppression and higher output signal-to-interference-plus-noise ratio (SINR) than competing methods in high-dimensional and impulsive noise regimes.
\end{abstract}
\vspace{-0.1cm}
\vspace{-0.3cm}

\section{Introduction}

Adaptive beamforming is a fundamental technique in wireless communications, radar, sonar, microphone arrays, and related sensing systems~\cite{brennan1976adaptive,krolik1996performance,gershman1995experimental,kaneda1986adaptive,godara1997application}. Its goal is to preserve the signal of interest while suppressing interference and noise. Among classical beamforming designs, the minimum variance distortionless response (MVDR) beamformer~\cite{capon1969high} is particularly attractive, as it minimizes the array output power subject to a distortionless response constraint in the look direction. In practice, however, the MVDR solution depends on the unknown inverse covariance matrix of the received data. Replacing this matrix with the sample covariance matrix (SCM) leads to the sample matrix inversion (SMI) beamformer, whose performance can deteriorate severely when the number of snapshots is limited~\cite{boroson1980sample,yang2018high} or when the data are contaminated by outliers and heavy-tailed noise~\cite{elkhalil2017fluctuations,ollila2003robust}.

Robust scatter estimators~\cite{maronna1976robust,huber2011robust,tyler1987distribution,kent1991redescending}, including Maronna's estimator \cite{maronna1976robust}, provide an effective way to reduce the influence of impulsive and non-Gaussian observations. However, robustness alone is insufficient in modern large-array regimes, where the number of sensors $N$ is comparable to the number of snapshots $n$. In such high-dimensional settings, finite-sample effects induce systematic distortions in both sample-based eigenvalues and eigenvectors, and these distortions are further amplified by matrix inversion in MVDR beamforming. Random matrix theory has therefore been widely used to analyze robust scatter estimators in the large-dimensional regime where $N,n\to\infty$ with $N/n$ bounded away from zero~\cite{couillet2015random,kammoun2017random,auguin2018large}. A common remedy is diagonal loading, which regularizes a robust scatter estimator by forming a linear combination with a scaled identity matrix~\cite{auguin2018large,couillet2014large,ollila2020shrinking}. Although diagonal loading improves numerical stability and mitigates sampling noise, it acts as a generic regularizer and does not directly exploit the low-rank signal-plus-interference structure that naturally arises in array covariance matrices. As shown later in this paper, such generic regularization can remain suboptimal for high-dimensional MVDR beamforming.


In~\cite{yang2018high}, a random-matrix-optimized MVDR beamformer was developed for high-dimensional spiked covariance models, where the population covariance matrix is represented as a low-rank perturbation of the noise covariance~\cite{donoho2018optimal,bai2012estimation,benaych2011eigenvalues}. By exploiting this structure, the method constructs a precision matrix estimator, i.e., an inverse covariance estimator, tailored to the MVDR output power. However, since it relies on the SCM, it remains sensitive to impulsive noise and heavy-tailed observations.

This paper aims to extend the random-matrix-optimized MVDR framework in~\cite{yang2018high} to impulsive and heavy-tailed noise environments. The results in~\cite{couillet2015robust} enable the analysis of robust spiked random matrices in the large-dimensional regime, where $N$ and $n$ grow at the same rate. Building on these tools, we propose a robust high-dimensional MVDR beamformer that combines Maronna's robust scatter estimator with spiked covariance techniques from random matrix theory. The proposed method is designed to mitigate two sources of performance degradation simultaneously: high-dimensional spectral distortion and non-Gaussian snapshot contamination. Numerical simulations under elliptically distributed noise show that the proposed beamformer remains stable and outperforms competing beamformers over a broad range of signal-to-noise ratios.

\vspace{-0.35cm}
\section{Background}

\subsection{Signal model and optimal MVDR beamforming}

We consider a uniform linear array with $N$ sensors receiving $m<N$ narrow-band sources. At snapshot $j\in\{1,\ldots,n\}$, the array observation is modeled as
{
\setlength{\abovedisplayskip}{2pt}
\setlength{\belowdisplayskip}{2pt}
\setlength{\abovedisplayshortskip}{1pt}
\setlength{\belowdisplayshortskip}{1pt}
\vspace{-0.05cm}
\begin{equation}
{\bm x}(j)
=
\sum_{i=1}^{m}\sqrt{p_i}{\bf a}(\theta_i)z_i(j)
+
{\bf n}(j),
\label{eq:robust_signal}
\vspace{-0.1cm}
\end{equation}
where ${\bf a}(\theta_i)\in\mathbb{C}^N$ denotes the unit-norm steering vector associated with direction $\theta_i$, $p_i$ is the corresponding source power, and $z_i(j)$ is an independent zero-mean unit-variance complex Gaussian signal. The first source is the signal of interest (SoI), while the remaining sources are regarded as interferers. The heavy-tailed noise is modeled as
${\bf n}(j)=\sigma\sqrt{\tau_j}{\bm o}(j)$, where ${\bm o}(j)$ is zero-mean and unitarily invariant with $\|{\bm o}(j)\|^2=N$, $\sigma$ denotes the noise level, and $\tau_j>0$ is a texture variable. The corresponding covariance matrix is
\vspace{-0.05cm}
\begin{equation}
{\bf C}_N
=
\sum_{i=1}^{m}p_i{\bf a}(\theta_i){\bf a}^H(\theta_i)
+
\sigma^2{\bf I}_N .
\label{eq:C_N}
\end{equation}

The classical MVDR beamformer ${\bf h}_{\rm MVDR}$ \cite{capon1969high} minimizes the output power
$P({\bf h})={\bf h}^H{\bf C}_N{\bf h}$ under a distortionless response constraint toward the SoI, namely
\begin{equation}
\min_{{\bf h}\in\mathbb{C}^N}
{\bf h}^H{\bf C}_N{\bf h},
\quad
\mathrm{s.t.}\ 
{\bf h}^H{\bf a}(\theta_1)=1 .
\vspace{-0.1cm}
\end{equation}
The resulting optimal beamformer is 
\vspace{-0.1cm}
\begin{align}
{\bf h}_{\rm MVDR}
=
\frac{{\bf C}_N^{-1}{\bf a}(\theta_1)}
{{\bf a}^H(\theta_1){\bf C}_N^{-1}{\bf a}(\theta_1)}
\end{align}
and its output power is $P({\bf h}_{\rm MVDR})
=
\frac{1}
{{\bf a}^H(\theta_1){\bf C}_N^{-1}{\bf a}(\theta_1)} .$
In the sequel, we use the normalized output power
$\rho({\bf h})=P({\bf h})/\sigma^2$ for notational convenience.
}

\subsection{Sample-based implementation of MVDR beamforming with robust scatter estimator}
\label{sec:Sample}
{
\setlength{\abovedisplayskip}{1pt}
\setlength{\belowdisplayskip}{1pt}
\setlength{\abovedisplayshortskip}{0pt}
\setlength{\belowdisplayshortskip}{0pt}

In practice, the population inverse covariance matrix ${\bf C}_N^{-1}$ is unknown and must be replaced by an estimate $\hat{\bf C}_N^{-1}$. This leads to the plug-in MVDR beamformer
\vspace{-0.1cm}
\begin{align}
\hat{\bf h}_{\rm MVDR}
=
\frac{\hat{\bf C}_N^{-1}{\bf a}(\theta_1)}
{{\bf a}^H(\theta_1)\hat{\bf C}_N^{-1}{\bf a}(\theta_1)} .
\label{eq:hest}
\end{align}
The resulting normalized total output power depends on the quality of the inverse covariance estimate and is given by
\begin{align}
\rho(\hat{\bf h}_{\rm MVDR})
=
\frac{1}{\sigma^2}
\frac{
{\bf a}^H(\theta_1)\hat{\bf C}_N^{-1}{\bf C}_N
\hat{\bf C}_N^{-1}{\bf a}(\theta_1)
}{
\left({\bf a}^H(\theta_1)\hat{\bf C}_N^{-1}{\bf a}(\theta_1)\right)^2
}.
\label{eq:L}
\end{align}
This quantity coincides with the oracle value
$\rho_{\rm min}=\rho({\bf h}_{\rm MVDR})$ only when
$\hat{\bf C}_N^{-1}={\bf C}_N^{-1}$. Otherwise, it quantifies the performance loss caused by imperfect inverse covariance estimation.

A classical approach is to estimate the covariance matrix by the SCM,
${\bf S}_N=n^{-1}\sum_{j=1}^n{\bf x}(j){\bf x}^H(j)$.
The beamformer obtained by replacing $\hat{\bf C}_N^{-1}$ with ${\bf S}_N^{-1}$ in \eqref{eq:hest} is commonly referred to as the sample matrix inversion (SMI) beamformer. However, the SMI beamformer may suffer from a substantial increase in normalized total output power relative to the theoretical optimum $\rho_{\rm min}$, especially in the presence of heavy-tailed noise or when the sensor number $N$ and the sample size $n$ are of the same order.

Robust scatter matrix estimators are widely used to mitigate the impact of outlier-contaminated or heavy-tailed observations. In this paper, we consider Maronna's robust scatter estimator
$\hat{\bf C}_{\rm R}$ \cite{maronna1976robust}, defined as the solution to
\begin{equation}
{\bf Z}
=
\frac{1}{n}\sum_{j=1}^{n}
u\left(
\frac{1}{N}{\bm x}^H(j){\bf Z}^{-1}{\bm x}(j)
\right)
{\bm x}(j){\bm x}^H(j),
\label{eq:robust_scatter}
\end{equation}
where $u$ is a nonnegative, nonincreasing, bounded, and continuous function on $\mathbb{R}^+$. Moreover,
$\phi(x)=xu(x)$ is assumed to be increasing and bounded, with
$\phi_\infty=\lim_{x\to\infty}\phi(x)>1$. In this work, we adopt
\vspace{-0.2cm}
\begin{equation}
u(x)=\frac{1+\alpha}{\alpha+x},
\qquad
\alpha>0,
\vspace{0.1cm}
\end{equation}
in which, for a proper choice of $\alpha$, this estimator corresponds to the maximum-likelihood scatter estimator under multivariate Student-distributed observations.

Although $\hat{\bf C}_{\rm R}$ improves robustness against heavy-tailed samples, it does not fully eliminate finite-sample spectral distortions in high-dimensional regimes, particularly when $N$ and $n$ are comparable. To address this issue, we exploit prior knowledge of the low-rank spiked structure of ${\bf C}_N$ and develop an optimized robust precision matrix estimator
$\hat{\bf C}_{\rm R,MVDR}^{-1}$. The resulting optimized beamformer
$\hat{\bf h}_{\rm R,MVDRopt}$ is designed to minimize \eqref{eq:L}, while improving robustness to both heavy-tailed noise and high-dimensional finite-sample effects.
}

\section{Optimized Robust High-Dimensional Beamformer Design}
\label{sec:proposed_method}
{
\setlength{\abovedisplayskip}{0pt}
\setlength{\belowdisplayskip}{0pt}
\setlength{\abovedisplayshortskip}{0pt}
\setlength{\belowdisplayshortskip}{0pt}

\subsection{Robust high-dimensional MVDR beamforming based on spiked covariance models}

We aim to construct a sample-based precision matrix estimator that is optimized for MVDR beamforming. Under the spiked covariance model~\cite{yang2018high,couillet2015robust}, the population covariance matrix ${\bf C}_N$ can be expressed as
\vspace{-0.2cm}
\begin{align}
{\bf C}_N
=
\sigma^2
\left(
{\bf I}_N+\sum_{i=1}^{m}t_i{\bf v}_i{\bf v}_i^H
\right),
\label{eq:CNhat_spk}
\end{align}
where $t_i>0$ denotes the strength of the $i$th spike and ${\bf v}_i$ is the corresponding population eigenvector.

Let the eigendecomposition of Maronna's robust scatter estimator be
$\hat{\bf C}_{\rm R}
=
\sum_{i=1}^{N}\lambda_i{\bf u}_i{\bf u}_i^H$,
with eigenvalues arranged in decreasing order. We consider precision matrix estimators of the form
\begin{align}
\hat{\bf C}_N^{-1}(\hat{\bf C}_{\rm R})
=
\sum_{i=1}^{N}\eta_i{\bf u}_i{\bf u}_i^H,
\label{eq:GeneralForm}
\end{align}
where the shrinkage coefficients $\eta_i>0$ are to be designed. Since only the largest $m$ eigenvalues are associated with the signal spikes, we hard-clip the remaining $N-m$ noise eigenvalues by setting
$\eta_{m+1}=\cdots=\eta_N=1/\sigma^2$. Defining $w_i=\sigma^2\eta_i-1$, we obtain
\vspace{-0.2cm}
\begin{align}
\hat{\bf C}_N^{-1}(\hat{\bf C}_{\rm R})
=
\frac{1}{\sigma^2}
\left(
{\bf I}_N+\sum_{i=1}^{m}w_i{\bf u}_i{\bf u}_i^H
\right).
\label{eq:SNhat_spk}
\end{align}

Substituting \eqref{eq:SNhat_spk} and \eqref{eq:CNhat_spk} into \eqref{eq:L}, the normalized total output power becomes a function of
${\bm w}=[w_1,\ldots,w_m]^T$, denoted by $\rho({\bm w})$. The explicit expression of $\rho({\bm w})$ is given in \eqref{eq:L_w2} at the top of the next page. The MVDR-oriented shrinkage design is therefore formulated as
\vspace{-0.2cm}
\begin{align}
{\bm w}^*
=
\argmin_{{\bm w}\in\mathcal{L}^m}
\rho({\bm w}),
\label{eq:L_w}
\end{align}
where
\footnote{The bounded constraint ${\bm w}\in\mathcal{L}^m$ is introduced as a technical condition for establishing the uniform convergence result in Theorem~\ref{THRM:DET}.}
$\mathcal{L}^m=[-1+\xi,q)^m$ for a small $\xi>0$ and a sufficiently large $q>0$. 
\begin{figure*}[t]
\begin{align}
\rho({\bm w})
=
\frac{
{\bf a}^H(\theta_1)
\left({\bf I}_N+\sum_{i=1}^{m}w_i{\bf u}_i{\bf u}_i^H\right)
\left({\bf I}_N+\sum_{j=1}^{m}t_j{\bf v}_j{\bf v}_j^H\right)
\left({\bf I}_N+\sum_{h=1}^{m}w_h{\bf u}_h{\bf u}_h^H\right)
{\bf a}(\theta_1)
}{
\left[
{\bf a}^H(\theta_1)
\left({\bf I}_N+\sum_{l=1}^{m}w_l{\bf u}_l{\bf u}_l^H\right)
{\bf a}(\theta_1)
\right]^2
}.
\label{eq:L_w2}
\end{align}
\hrule
\end{figure*}

Directly solving \eqref{eq:L_w} is challenging, and even if such an optimal ${\bm w}^*$ were obtained, this could not be used in practice, since it involves unobservable quantities (i.e., $t_i$ and ${\bf v}_i$), which are properties of ${\bf C}_N$. Inspired by~\cite{yang2018high}, we address this difficulty using asymptotic tools from random matrix theory. Specifically, we first derive an asymptotic deterministic equivalent $\bar{\rho}({\bm w})$ of $\rho({\bm w})$ as $N,n\to\infty$, and characterize the oracle optimizer $\bar{\bm w}^*$ that minimizes $\bar{\rho}({\bm w})$. We then develop a fully sample-based consistent estimator $\hat{\bm w}^*$, which leads to the proposed robust MVDR precision matrix estimator
$\hat{\bf C}_{\rm R,MVDR}^{-1}$ and the corresponding optimized beamformer
$\hat{\bf h}_{\rm R,MVDRopt}$. 

\subsection[Robust deterministic equivalent]{Robust deterministic equivalent $\bar{\rho}({\bm w})$ and the optimal ${\bm w}^*$}
For our asymptotic analysis, we assume the
following:
\vspace{-0.2cm}
\begin{assumption} \label{assump}
\hfill
\vspace{-0.8em}
\begin{enumerate}[a.]
\item $\tau_1, \ldots, \tau_n \in (0,\infty)$ are random scalars such that
$\nu_n \triangleq \frac{1}{n}\sum_{i=1}^{n}\delta_{\tau_i} \to \nu$
weakly, almost surely, where $\int t \, \nu(dt) = 1$;
\item As $N,n\rightarrow\infty$, $N/n=c_N\rightarrow c$ for a certain $c>0$;\label{assump_a}
\item The number of spikes $m$ is fixed, independently of $N$ and $n$.\label{assump_b}
\end{enumerate}
\vspace{-0.8em}
\end{assumption}

To derive the asymptotic deterministic equivalent $\bar{\rho}({\bm w})$ of $\rho({\bm w})$ under Assumption \ref{assump}, we first introduce some notation. Following the same definitions as in \cite{couillet2015robust},
define
\begin{equation}
v(x)=u(g^{-1}(x)),
\qquad
\psi(x)=xv(x)
\end{equation}
where $g^{-1}$ denotes the inverse function of $g(x)=\frac{x}{1-c\phi(x)}$,
and let $\gamma>0$ be the solution of
\begin{equation}
1=
\int
\frac{\psi(\tau\gamma)}
{1+c\,\psi(\tau\gamma)}
\nu(d\tau).
\label{eq:gamma_known_nu}
\end{equation}

Let us first consider the noise-only setting in which
$p_1=\cdots=p_m=0$. Then, by~\cite[Theorem 1]{couillet2015robust} and~\cite{silverstein1995analysis},
the empirical spectral distribution $\mu_n \triangleq \frac{1}{N}\sum_{i=1}^{N}\delta_{\lambda_i}$
converges weakly, almost surely, to a limiting distribution $\mu$. The
measure $\mu$ admits a density on $\mathbb{R}$ with bounded support satisfying
$\operatorname{Supp}(\mu)\subset\mathbb{R}^+$. Denote
\[
S_\mu^- \triangleq \inf(\operatorname{Supp}(\mu)),
\qquad
S_\mu^+ \triangleq \sup(\operatorname{Supp}(\mu)).
\]
For $x\in\mathbb{R}\setminus [S_\mu^-,S_\mu^+]$, define $\delta(x)$ as the
unique real solution of
\begin{equation}
\delta(x)
=
c\left(
-x+
\int
\frac{\tau v(\tau\gamma)}
{1+\delta(x)\tau v(\tau\gamma)}
\,\nu(d\tau)
\right)^{-1}.
\label{eq:delta_known_nu}
\end{equation}
Coming back to generic values for $p_1,\ldots,p_m$, from \cite[Theorem 2]{couillet2015robust}, we have the following convergence result,
\begin{align} \label{eq;quadForm}
\left|{\bf v}_i^H{\bf u}_j{\bf u}_j^H{\bf v}_i-s_i\delta_{ij}\right|\stackrel{\rm a.s.}\longrightarrow0\,,  \quad \quad i,j=1,\ldots,m
\end{align}
with $\delta_{ij}$  the Kronecker-delta function, where 
\small
  \begin{align}\nonumber
  s_i \!=\!
  \frac{
  \displaystyle \int\!\!
  \tfrac{v(\tau\gamma)}
  {1+\delta({\lambda_i})\tau v(\tau\gamma)}
  \nu(d\tau)
  \left(
  1\!-\!\frac{1}{c}
  \int\!\!
  \tfrac{\delta({\lambda}_i)^2\tau^2 v(\tau\gamma)^2}
  {\left(1+\delta({\lambda}_i)\tau v(\tau\gamma)\right)^2}
  \nu(d\tau)
  \right)
  }
  {
  \displaystyle \int
  \tfrac{v(\tau\gamma)}
  {\left(1+\delta({\lambda}_i)\tau v(\tau\gamma)\right)^2}
  \nu(d\tau)
  } .
  \end{align}
  \normalsize
Define the deterministic quantities $k_i = {\bf a}^H(\theta_1){\bf v}_i{\bf v}_i^H{\bf a}(\theta_1)$, $i=1,\ldots,m$. We have the following result. 
\begin{theorem}[Robust Deterministic Equivalent] \label{THRM:DET}
Let Assumption \ref{assump} hold. As $N,n\to\infty$, $\sup_{\bm w\in\mathcal{L}^m}
\left|
\rho(\bm w)-\bar\rho(\bm w)
\right|
\xrightarrow{\rm a.s.}0$
where 
\begin{equation}
\bar\rho(\bm w)
=
\frac{
\bm w^T\bm B_R\bm w+2\bm w^T\bm d_R+a_R
}{
(1+\bm w^T\bm e_R)^2
}
\label{eq:bar_rho_R}
\end{equation}
with 
\vspace{-0.2cm}
\begin{align*}
\bm B_R
&=
\diag
\left\{
s_1 k_1(1+t_1 s_1), \ldots, s_m k_m(1+t_m s_m)
\right\}, \\ \bm d_R
&=
\left[
s_1 k_1(1+t_1), \ldots, s_m k_m(1+t_m)
\right], \\
\bm e_R
&=
\left[
s_1 k_1, \ldots, s_m k_m
\right]^T, ~~
a_R
=
1+\sum_{i=1}^{m}t_ik_i.
\end{align*}
\end{theorem}
\vspace{-0.2cm}
\emph{Proof:} See Appendix~\ref{app:proof-det}.

We now seek the value of $\bar{\bm w}^*=[\bar{w}_1^*,\ldots,\bar{w}_m^*]^T$ that minimizes $\bar{\rho}({\bm w})$ in (\ref{eq:bar_rho_R}). This is given by the following result.
\begin{theorem}[MVDR-Optimal Weights] \label{THRM:DET_OPT}
Under the setting of Theorem~\ref{THRM:DET}, 
\vspace{-0.4cm}
\begin{align}
\bar{\bm w}^*
=
\argmin_{\bm w\in\mathcal{L}^m}\bar\rho(\bm w)
\end{align}
where, for $i=1,\ldots,m$,
\vspace{-0.6cm} 
\begin{align}
\bar{w}_{i}^*
=
\frac{
\chi-t_i
}{
1+t_is_i
}
~~{\rm with}~~\chi
=
\frac{
\displaystyle
\sum_{j=1}^{m}
\frac{t_jk_j(1-s_j)}
{1+t_js_j}
}{
\displaystyle
\sum_{j=1}^{m}
\frac{k_j(1-s_j)}
{1+t_js_j}
}.
\end{align}
\end{theorem}
\emph{Proof:} See Appendix~\ref{app:proof-det-opt}.

\subsection[Estimated optimal weights and proposed algorithm]{Estimated optimal weights $\hat{\bm w}^*$ and proposed algorithm}
The optimal weights in Theorem~\ref{THRM:DET_OPT} depend on the unobservable quantities $t_i$, $s_i$, and $k_i$, and are therefore not directly applicable in practice. We next provide consistent sample-based estimators of these weights using the leading eigenvalues $\lambda_i$ and eigenvectors ${\bf u}_i$ of the robust scatter matrix $\hat{\bf C}_{\rm R}$, for $i=1,\ldots,m$.

Let $\hat{\bf C}_{{\rm R},(j)}= \hat{\bf C}_{\rm R}
- \frac{1}{n} u\left(\frac{1}{N}{\bm x}^H(j)\hat{\bf C}_{\rm R}^{-1}{\bm x}(j)\right)
{\bm x}(j){\bm x}^H(j)$
denote the
leave-one-out robust scatter estimator obtained by removing the $j$th sample. Define $\hat\gamma_n
=
\frac{1}{n}\sum_{j=1}^{n}
\frac{1}{N}\bm x^H(j)\hat{\bf C}_{{\rm R},(j)}^{-1}\bm x(j)$ and $\hat\tau_j
=
\frac{1}{\hat\gamma_n}
\frac{1}{N}\bm x^H(j)\hat{\bf C}_{{\rm R},(j)}^{-1}\bm x(j)$.
For $x\in(S_\mu^+,\infty)$, we denote by $\hat\delta(x)$ the unique negative solution to
\vspace{-0.2cm}
\begin{equation}
\hat\delta(x)
=
c_N
\left(
-x+
\frac{1}{n}\sum_{j=1}^{n}
\frac{
\hat\tau_j v(\hat\tau_j\hat\gamma_n)
}{
1+\hat\delta(x)\hat\tau_j v(\hat\tau_j\hat\gamma_n)
}
\right)^{-1}.
\label{eq:sample_delta}
\vspace{-0.2cm}
\end{equation}
For each robust spike eigenvalue $\lambda_i$, set
\vspace{-0.2cm}
\begin{equation}
\hat\delta_i=\hat\delta(\lambda_i),
\qquad
v_j=v(\hat\tau_j\hat\gamma_n).
\label{eq:sample_delta_v}
\vspace{-0.2cm}
\end{equation}
The following theorem gives a
fully sample-based version of the robust MVDR-optimal weights.
\begin{theorem}[Estimated Optimal Weights]\label{THRM:SAMPLE_BASED}
Under the setting of Theorem~\ref{THRM:DET}, for all large $n$ with probability one, we have, for $i=1,\ldots,m$, 
{
\setlength{\abovedisplayskip}{2pt}
\setlength{\belowdisplayskip}{0pt}
\setlength{\abovedisplayshortskip}{2pt}
\setlength{\belowdisplayshortskip}{0pt}
\begin{align}
\left|\hat{w}_i^*-\bar{w}_i^*\right|
\stackrel{\rm a.s.}\longrightarrow 0 .
\label{w_conv}
\end{align}
}
where $\hat{w}_{i}^*
=
\dfrac{
\hat\chi-\hat{t}_i
}{
1+\hat{t}_i\hat{s}_i
}$, in which
\begin{align*}
\hat\chi
&=
\frac{
\displaystyle
\sum_{j=1}^{m}
\frac{\hat{t}_j\hat{k}_j(1-\hat{s}_j)}
{1+\hat{t}_j\hat{s}_j}
}{
\displaystyle
\sum_{j=1}^{m}
\frac{\hat{k}_j(1-\hat{s}_j)}
{1+\hat{t}_j\hat{s}_j}
}, ~~
\hat t_i
=
-
\left(
\hat\delta_i
\frac{1}{N}
\sum_{j=1}^{n}
\frac{v_j}
{1+\hat\delta_i\hat\tau_j v_j}
\right)^{-1}, \\
\hat s_i
&=
\frac{
\displaystyle
\left(
\frac{1}{n}\sum_{j=1}^{n}
\frac{v_j}
{1+\hat\delta_i\hat\tau_jv_j}
\right)
\left[
1-\frac{1}{N}\sum_{j=1}^{n}
\frac{\hat\delta_i^2\hat\tau_j^2v_j^2}
{(1+\hat\delta_i\hat\tau_jv_j)^2}
\right]
}{
\displaystyle
\frac{1}{n}\sum_{j=1}^{n}
\frac{v_j}
{(1+\hat\delta_i\hat\tau_jv_j)^2}
}, \\
\hat k_i
&=
\frac{1}{\hat s_i}
{\bf a}^H(\theta_1){\bf u}_i{\bf u}_i^H{\bf a}(\theta_1).
\end{align*}
\end{theorem}
\emph{Proof:} See Appendix~\ref{app:proof-sample-based}.

This leads to our proposed robust high-dimensional MVDR beamformer construction summarized in Algorithm 1.
\begin{algorithm}[t]
\caption{Robust High-Dimensional MVDR Beamformer}
\begin{algorithmic}[1]
\STATE Compute $\hat{w}_i^*$, $i=1,\ldots,m$, according to Theorem~\ref{THRM:SAMPLE_BASED}.
\STATE Form
$\displaystyle
\hat{\mathbf{C}}_{\mathrm{R,MVDR}}^{-1}
=
\frac{1}{\sigma^2}
\left(
\mathbf{I}_N+
\sum_{i=1}^{m}
\hat{w}_i^*
\mathbf{u}_i\mathbf{u}_i^{H}
\right)
$ ($\frac{1}{\sigma^2}$ can be omitted if only the beamformer is needed).
\STATE Construct
$\displaystyle
\hat{\mathbf{h}}_{\mathrm{R,MVDRopt}}
=
\frac{
\hat{\mathbf{C}}_{\mathrm{R,MVDR}}^{-1}\mathbf{a}(\theta_1)
}{
\mathbf{a}^{H}(\theta_1)
\hat{\mathbf{C}}_{\mathrm{R,MVDR}}^{-1}
\mathbf{a}(\theta_1)
}
$.
\end{algorithmic}
\end{algorithm}
}

\section{Numerical Simulations}
{
\setlength{\abovedisplayskip}{2pt}
\setlength{\belowdisplayskip}{2pt}
\setlength{\abovedisplayshortskip}{0pt}
\setlength{\belowdisplayshortskip}{0pt}
For our simulations, we consider a uniform linear array with $N$ identical omnidirectional sensors placed at half-wavelength spacing. The SoI impinges from direction $\theta_1=0^\circ$, while five interferers impinge from directions $\theta_2=5^\circ,
\theta_3=10^\circ,
\theta_4=30^\circ,
\theta_5=50^\circ,
\theta_6=70^\circ$. 
The steering vector is ${\bf a}(\theta)=\frac{1}{\sqrt{N}}[1, e^{j\pi \sin(\theta)},\ldots,e^{j\pi \sin(\theta)(N-1)}]^T$.
The texture variables $\tau_j$ in
\eqref{eq:robust_signal} are generated from a normalized Student-type
texture model,
\[
\tau_j\stackrel{d}{=}\frac{\beta-2}{\beta}t_j^2,
\qquad t_j\sim t_\beta ,
\]
where $t_\beta$ denotes a standard Student-$t$ random variable with
$\beta$ degrees of freedom. The factor $(\beta-2)/\beta$ normalizes the
texture so that $\mathbb{E}[\tau_j]=1$ for $\beta>2$. Here $\beta$ controls the tail heaviness, with smaller $\beta$ producing more impulsive snapshots; we set $\beta=5$ throughout the baseline
simulations.  The function $u$ in $\hat{\bf C}_{\rm R}$ is
$u(x)=\frac{1+\alpha}{\alpha+x}$ with $\alpha=0.2$.
Each data point is averaged over $200$ independent Monte Carlo trials. 
}
\vspace{-0.25cm}
 \begin{figure}[H]
\centering
\includegraphics[width=0.8\linewidth]{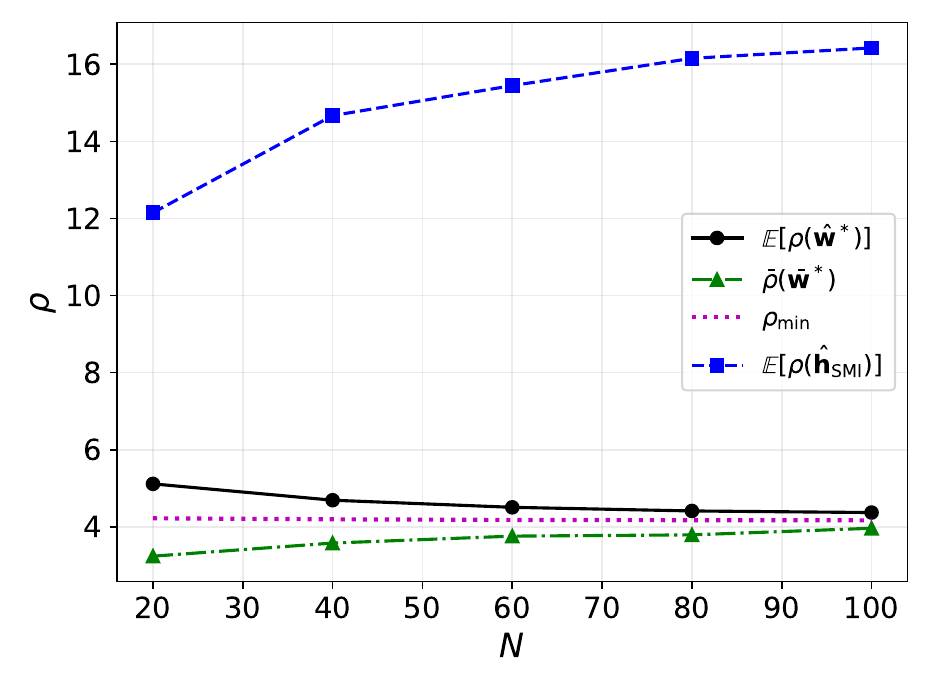}
\vspace{-0.5cm}
\caption{$\rho$-performance when $n=2N$.}
\label{fig:robust_rho}
\vspace{-0.25cm}
\end{figure}

\subsection[Output power convergence and the deterministic equivalent]{$\rho$-performance and the deterministic equivalent}
We first investigate the convergence of the proposed algorithm in terms of the normalized output power $\rho$. The signal-to-noise ratio (SNR) and interference-to-noise ratio (INR) are defined as SNR $=p_1/\sigma^2$ and INR $=p_i/\sigma^2$, with the same INR for all $i=2,\ldots,m$.

In Fig.~\ref{fig:robust_rho}, for SNR $=5$ dB, INR $=30$ dB, and $n=2N$, we compare the empirical expectation $\mathbb{E}[\rho(\hat{\bm w}^*)]$ obtained from Theorem~\ref{THRM:SAMPLE_BASED} with the deterministic equivalent $\bar{\rho}(\bar{\bm w}^*)$ given by Theorems~\ref{THRM:DET} and~\ref{THRM:DET_OPT}.
 Note that, as expected, $\mathbb{E}[ \rho (\hat{\bm w}^*)]$ converges to $\bar{\rho}(\bar{\bm w}^*)$ as $N$ and $n$ increased.  For additional comparison, we also plot the theoretical minimum $\rho_{\rm min}= 1 / (\sigma^2{\bf a}^H(\theta_1){\bf C}_N^{-1}{\bf a}(\theta_1))$ and the expectation $\mathbb{E} [\rho( \hat{\bf h}_{\rm SMI})]$ for the SMI beamformer. These results demonstrate that $\mathbb{E} [ \rho (\hat{\bm w}^*)]$  is close to $\rho_{\rm min}$, indicating that our proposed approach leads to near-optimal performance. On the other hand, $\mathbb{E} [\rho( \hat{\bf h}_{\rm SMI})]$ is larger than $\mathbb{E}[ \rho (\hat{\bm w}^*)]$ over the entire range of $N$.

\subsection{Beamformer performance and comparison with baselines}
{
\setlength{\abovedisplayskip}{2pt}
\setlength{\belowdisplayskip}{2pt}
\setlength{\abovedisplayshortskip}{0pt}
\setlength{\belowdisplayshortskip}{0pt}
We next compare the proposed beamformer $\hat{\bf h}_{\rm R,MVDRopt}$
with several representative alternatives in terms of beampattern and output
SINR. The beampattern measures the response to a source arriving from
direction $\kappa$ and is computed as
$\mathbb{E}\left[20\log_{10}\left|\hat{\bf h}^{H}{\bf a}(\kappa)\right|\right]$,
where the expectation is approximated by averaging over $200$ independent
Monte Carlo trials and $\kappa$ is swept from $-90^\circ$ to $90^\circ$.
Unless otherwise stated, we set INR $=30$ dB and $n=2N=200$.

The proposed method is compared with the oracle MVDR beamformer constructed
from the true covariance matrix ${\bf C}_N$, the SMI beamformer
$\hat{\bf h}_{\rm SMI}$, and two robust baselines. The first robust baseline
is the robust diagonal loading beamformer $\hat{\bf h}_{\rm RDL}$, which uses
$\hat{\bf C}_{\rm RDL}(\varphi)=(1-\varphi)\hat{\bf C}_{\rm R}
+\varphi{\bf I}_N$ with $\varphi\in[0,1]$. While many methods have been developed to specify the diagonal loading parameter $\varphi$~\cite{auguin2018large,ollila2020shrinking,couillet2014large}, here we compare with the empirically computed ``oracle'' solution, in which $\varphi$ is chosen to minimize the normalized total output power in (\ref{eq:L}). Note that this optimal solution is unobtainable in practice, but it provides an upper bound on the performance achievable with the diagonal loading method. The second is the robust
eigen-subspace beamformer $\hat{\bf h}_{\rm REigsub}$~\cite{chang1992performance,feldman1994projection}, defined by
\[
\hat{\bf h}_{\rm REigsub}
=
\frac{\hat{\bf C}_{\rm R}^{-1}{\bf a}_{\rm sub}}
{{\bf a}_{\rm sub}^{H}\hat{\bf C}_{\rm R}^{-1}{\bf a}_{\rm sub}},
\qquad
{\bf a}_{\rm sub}
=
\sum_{i=1}^{m}{\bf u}_i{\bf u}_i^H{\bf a}(\theta_1).
\]
It constructs a beamformer based on projecting ${\bf a}(\theta_1)$ onto the signal-plus-interference subspace estimated from $\hat{\bf C}_{\rm R}$. 
Fig.~\ref{fig:beampattern_all} shows the beampatterns at SNR $=5$ dB.
The proposed $\hat{\bf h}_{\rm R,MVDRopt}$ places clear nulls toward the
interferers while maintaining a stable response in the look direction. In
contrast, $\hat{\bf h}_{\rm SMI}$ exhibits a noticeably higher sidelobe and
noise floor, reflecting the instability of direct sample covariance inversion
under heavy-tailed observations and limited snapshots. The robust baselines
$\hat{\bf h}_{\rm RDL}$ and $\hat{\bf h}_{\rm REigsub}$ improve over SMI, but
their interference suppression is less pronounced than that of the proposed
$\hat{\bf h}_{\rm R,MVDRopt}$.
\vspace{-0.05cm}
\begin{figure}[!h]
\centering
\subfigure[Beampatterns (SNR $=5$ dB)]{
\includegraphics[width=0.8\linewidth]{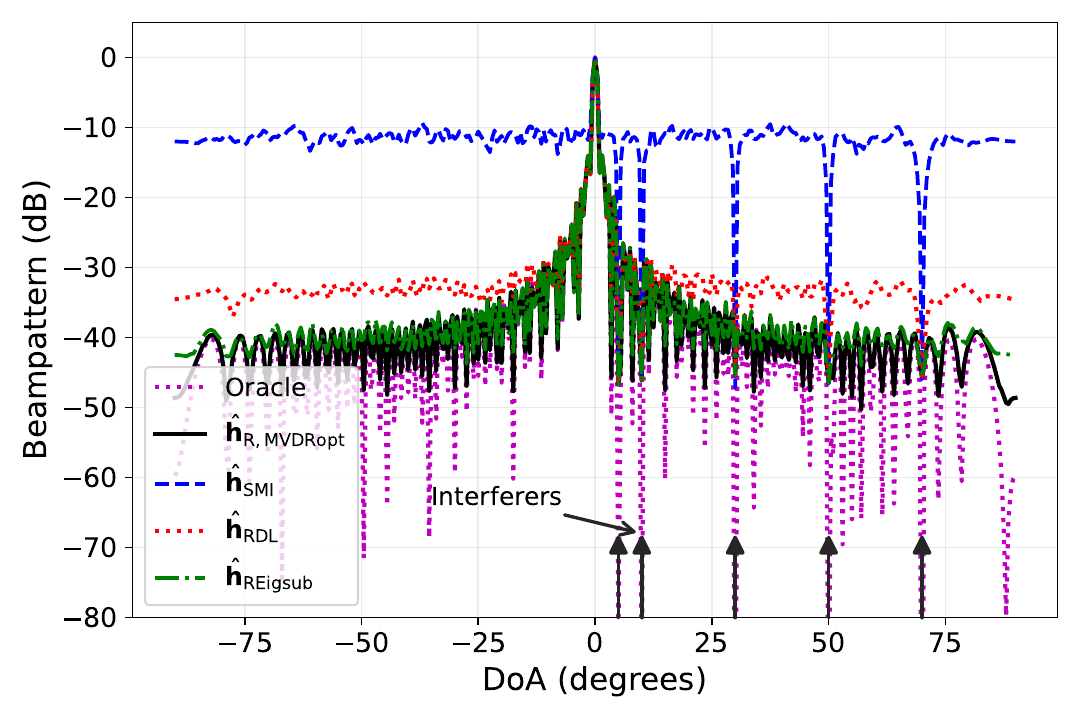}
\label{fig:beampattern_all}
}
\subfigure[Output SINR]{
\includegraphics[width=0.8\linewidth]{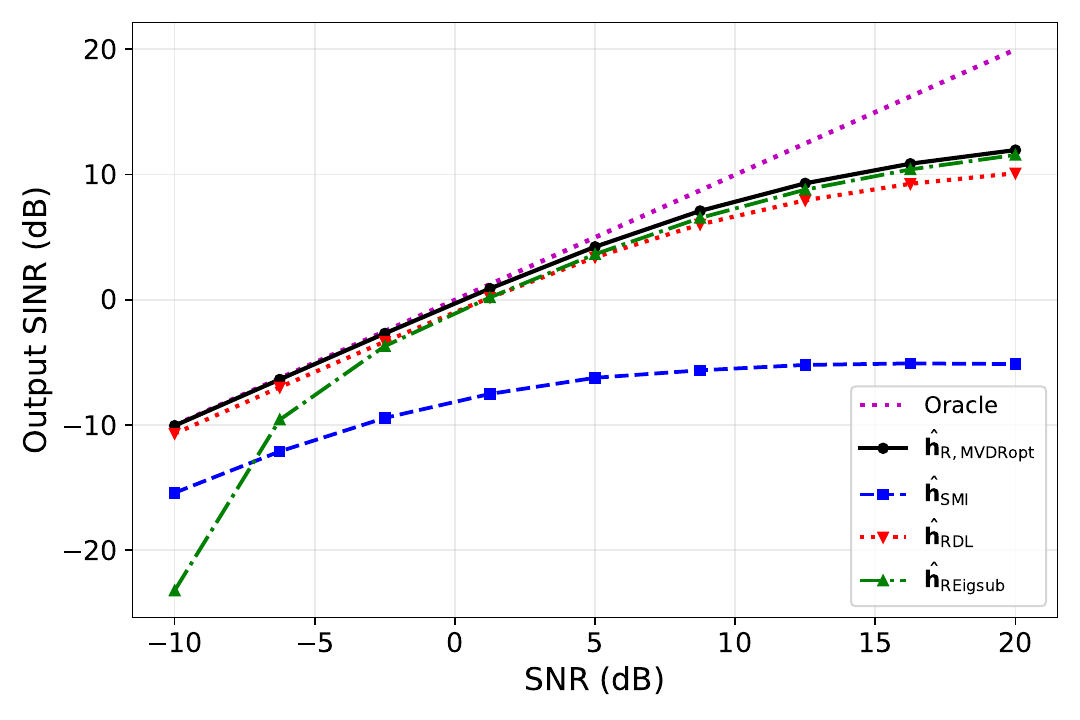}
\label{fig:sinr}}
\vspace{-0.4cm}
\caption{Beampatterns and SINR performance.}
\label{fig:performance}
\vspace{-0.2cm}
\end{figure}

The SINR results in Fig.~\ref{fig:sinr} lead to the same conclusion. $\hat{\bf h}_{\rm R,MVDRopt}$ consistently achieves the highest output SINR among the
sample-based methods and remains the closest to the oracle MVDR benchmark. Its performance is comparable to $\hat{\bf h}_{\rm RDL}$ at low SNR and to $\hat{\bf h}_{\rm REigsub}$ at high SNR.
}

\section{Conclusion}
This paper proposed a robust high-dimensional MVDR beamforming method for array observations corrupted by heavy-tailed noise. Rather than estimating the covariance matrix in a generic manner, the proposed approach directly constructs an MVDR-oriented precision matrix estimator by combining Maronna's robust scatter estimator with spiked covariance modeling. Using tools from random matrix theory, particularly robust spiked covariance models, we derived a deterministic equivalent of the MVDR output power, which leads to closed-form asymptotically optimal shrinkage eigenvalues and a fully sample-based implementation. Numerical simulations highlight the benefit of jointly exploiting robustness and high-dimensional spectral structure in adaptive beamforming. Future work will consider extensions to scenarios with steering vector mismatch.

\bibliographystyle{IEEEtran}
\bibliography{cited_2}

\clearpage
\appendix
\section{Proofs of Theorem~\ref{THRM:DET}}\label{app:proof-det}
The proof follows the deterministic-equivalent argument used for the
SCM-based MVDR estimator in \cite[Appendix A]{yang2018high}, with the SCM
eigenvector-overlap limits replaced by their robust spiked random
matrix counterparts in \cite[Ths. 2 and 3]{couillet2015robust}.

Define the robust sample rank-one projectors
\[
\hat{\bm P}_i
=
{\bf u}_i{\bf u}_i^{H},
\qquad i=1,\ldots,m,
\]
and 
\[
\hat{\bm Q}_R(\bm w)
=
\bm I_N+\sum_{i=1}^{m}w_i\hat{\bm P}_i.
\]
Since the robust spikes are separated from the robust noise
bulk, the robust spiked eigenvector results imply, for
$i,j,\ell\in\{1,\ldots,m\}$,
\begin{align}
{\bf a}^H(\theta_1)\hat{\bm P}_i{\bf a}(\theta_1)
&\xrightarrow{\rm a.s.}
s_ik_i,
\label{eq:proof_overlap_a}\\
{\bf a}^H(\theta_1)\hat{\bm P}_i
{\bf v}_j{\bf v}_j^H{\bf a}(\theta_1)
&\xrightarrow{\rm a.s.}
s_ik_i\,\delta_{ij},
\label{eq:proof_overlap_cross}\\
{\bf a}^H(\theta_1)\hat{\bm P}_i
{\bf v}_j{\bf v}_j^H
\hat{\bm P}_\ell{\bf a}(\theta_1)
&\xrightarrow{\rm a.s.}
s_i^2 k_i\,
\delta_{ij}\delta_{i\ell}.
\label{eq:proof_overlap_quad}
\end{align}
These are the robust analogues of the projection limits used in the
SCM proof. Equation \eqref{eq:proof_overlap_a} gives the asymptotic
projection of the steering vector on the robust sample spike
eigenvectors; \eqref{eq:proof_overlap_cross} and
\eqref{eq:proof_overlap_quad} are the corresponding first- and
second-order projections through the population spike directions.

Write the finite-sample output power in the form
\[
\rho(\bm w)
=
\frac{N(\bm w)}{D(\bm w)}
\]
with
\begin{align}
N(\bm w)
&=
{\bf a}^H(\theta_1)
\hat{\bm Q}_R(\bm w)
\left( {\bf I}_N+ \sum_{j=1}^m t_j{\bf v}_j{\bf v}_j^H \right)
\hat{\bm Q}_R(\bm w)
{\bf a}(\theta_1),\\
D(\bm w)
&=
\left|
{\bf a}^H(\theta_1)\hat{\bm Q}_R(\bm w){\bf a}(\theta_1)
\right|^2.
\end{align}
We first compute the deterministic limit of the denominator. From
\eqref{eq:proof_overlap_a},
\[
\bm a^H(\theta_1)\hat{\bm Q}_R(\bm w)\bm a(\theta_1)
=
1+\sum_{i=1}^{m}
w_i\bm a^H(\theta_1)\hat{\bm P}_i\bm a(\theta_1)
\xrightarrow{\rm a.s.}
1+\bm w^T\bm e_R.
\]
Thus, for every fixed $\bm w$,
\[
D(\bm w)
\xrightarrow{\rm a.s.}
(1+\bm w^T\bm e_R)^2.
\]

Next expand the numerator. The identity part gives
\begin{align}
\bm a^H(\theta_1)\hat{\bm Q}_R^2(\bm w)\bm a(\theta_1)
&=
1+
\sum_{i=1}^{m}(2w_i+w_i^2)
\bm a^H(\theta_1)\hat{\bm P}_i\bm a(\theta_1) \nonumber\\
&\xrightarrow{\rm a.s.}
1+
\sum_{i=1}^{m}
(2w_i+w_i^2)s_ik_i .
\label{eq:proof_identity_part}
\end{align}
For the spike part, \eqref{eq:proof_overlap_cross} and
\eqref{eq:proof_overlap_quad} yield
\begin{align}
&\bm a^H(\theta_1)
\hat{\bm Q}_R(\bm w)
{\bf v}_j{\bf v}_j^H
\hat{\bm Q}_R(\bm w)
\bm a(\theta_1) \nonumber\\
&\quad =
\left|
{\bf v}_j^H\bm a(\theta_1)+
\sum_{i=1}^{m}
w_i{\bf v}_j^H\hat{\bm P}_i\bm a(\theta_1)
\right|^2 \nonumber\\
&\quad \xrightarrow{\rm a.s.}
k_j\left(1+w_j s_j\right)^2 .
\label{eq:proof_spike_part}
\end{align}
Combining \eqref{eq:proof_identity_part} and
\eqref{eq:proof_spike_part}, for every fixed $\bm w$,
\begin{align}
N(\bm w)
&\xrightarrow{\rm a.s.}
1+\sum_{i=1}^{m}t_ik_i+
2\sum_{i=1}^{m}
w_i s_ik_i(1+t_i) \nonumber\\
&\quad+
\sum_{i=1}^{m}
w_i^2s_ik_i
\left(1+t_is_i\right) \nonumber\\
&=
\bm w^T\bm B_R\bm w
+2\bm w^T\bm d_R+a_R .
\label{eq:proof_num_limit}
\end{align}
Together with the denominator limit, this proves
$|\rho(\bm w)-\bar\rho(\bm w)|\to0$ almost surely for each fixed
$\bm w$.

It remains to upgrade pointwise convergence to uniform convergence on
$\mathcal L_m$. Both $N(\bm w)$ and $D(\bm w)$ are polynomials in
the finite-dimensional vector $\bm w$ of degree at most two. Their
random coefficients are precisely finite linear combinations of the
projection quantities in
\eqref{eq:proof_overlap_a}--\eqref{eq:proof_overlap_quad}; hence all
these coefficients converge almost surely to the deterministic
coefficients appearing in \eqref{eq:bar_rho_R}. Since
$\mathcal L_m$ is compact, the monomials
$1,w_i,w_iw_j$ are uniformly bounded on $\mathcal L_m$, and therefore
\begin{align}\label{eq:nominator}
\sup_{\bm w\in\mathcal L_m}
\left|
N(\bm w)-
\left(
\bm w^T\bm B_R\bm w+2\bm w^T\bm d_R+a_R
\right)
\right|
&\xrightarrow{\rm a.s.}0,\\ \label{eq:denominator}
\sup_{\bm w\in\mathcal L_m}
\left|
D(\bm w)-
(1+\bm w^T\bm e_R)^2
\right|
&\xrightarrow{\rm a.s.}0.
\end{align}

Finally, because $|1+\bm w^T\bm e_R|\ge\epsilon$ on $\mathcal L_m$, the
deterministic denominator
\[
\bar D(\bm w)=(1+\bm w^T\bm e_R)^2
\]
satisfies
\[
\inf_{\bm w\in\mathcal L_m}\bar D(\bm w)\ge\epsilon^2.
\]
The uniform denominator convergence implies that, for all sufficiently
large $N,n$ almost surely,
\[
\sup_{\bm w\in\mathcal L_m}|D(\bm w)-\bar D(\bm w)|
\le \frac{\epsilon^2}{2}.
\]
Therefore, for every $\bm w\in\mathcal L_m$ and all sufficiently large
$N,n$,
\[
D(\bm w)
\ge
\bar D(\bm w)-|D(\bm w)-\bar D(\bm w)|
\ge
\epsilon^2-\frac{\epsilon^2}{2}
=
\frac{\epsilon^2}{2}.
\]
This proves that the random denominator is also uniformly bounded away
from zero. Let
$\bar\rho(\bm w)=\bar N(\bm w)/\bar D(\bm w)$. Since
$\mathcal L_m$ is compact, $\bar N(\bm w)$ and $\bar D(\bm w)$
are uniformly bounded on $\mathcal L_m$. For all large $N,n$,
\begin{align}
\left|
\frac{N(\bm w)}{D(\bm w)}
-
\frac{\bar N(\bm w)}{\bar D(\bm w)}
\right|
&\le
\frac{|N(\bm w)-\bar N(\bm w)|}{D(\bm w)}
+
|\bar N(\bm w)|
\frac{|D(\bm w)-\bar D(\bm w)|}
{D(\bm w)\bar D(\bm w)} \nonumber\\
&\le
C
\left(
|N(\bm w)-\bar N(\bm w)|
+
|D(\bm w)-\bar D(\bm w)|
\right),
\end{align}
where $C<\infty$ is deterministic and independent of $\bm w$.
Taking the supremum over $\mathcal L_m$ and using the two uniform
polynomial convergences (\ref{eq:nominator}) and (\ref{eq:denominator}) gives
\[
\sup_{\bm w\in\mathcal L_m}
\left|
\rho(\bm w)-\bar\rho(\bm w)
\right|
\xrightarrow{\rm a.s.}0.
\]
This completes the proof.

\section{Proof of Theorem~\ref{THRM:DET_OPT}}\label{app:proof-det-opt}
The robust deterministic equivalent in \eqref{eq:bar_rho_R} can be
written as
\[
\bar\rho(\bm w)
=
\frac{q(\bm w)}{r^2(\bm w)}
\]
where
\begin{align}
q(\bm w)
&=
\bm w^T\bm B_R\bm w+2\bm w^T\bm d_R+a_R,\\
r(\bm w)
&=
1+\bm w^T\bm e_R.
\end{align}

For an interior optimum with $r(\bm w)\ne0$, the gradient must
vanish. Since
\[
\nabla q(\bm w)=2(\bm B_R\bm w+\bm d_R),
\qquad
\nabla r(\bm w)=\bm e_R,
\]
we obtain
\begin{align}
\nabla\bar\rho(\bm w)
&=
\frac{
2(\bm B_R\bm w+\bm d_R)r^2(\bm w)
-2q(\bm w)r(\bm w)\bm e_R
}{
r^4(\bm w)
} \nonumber\\
&=
\frac{2}{r^3(\bm w)}
\left[
r(\bm w)(\bm B_R\bm w+\bm d_R)
-q(\bm w)\bm e_R
\right].
\label{eq:grad_bar_rho_R}
\end{align}
Thus any interior stationary point satisfies
\begin{equation}
\bm B_R\bm w+\bm d_R
=
l\bm e_R,
\qquad
l
:=
\frac{q(\bm w)}{r(\bm w)}.
\label{eq:stationary_R}
\end{equation}
Because $\bm B_R$ is nonsingular, \eqref{eq:stationary_R} gives
\[
\bm w=\bm B_R^{-1}(l\bm e_R-\bm d_R).
\]
It remains to determine $l$. Let
\[
E_R=\bm e_R^T\bm B_R^{-1}\bm e_R,\quad
F_R=\bm e_R^T\bm B_R^{-1}\bm d_R,\quad
G_R=\bm d_R^T\bm B_R^{-1}\bm d_R.
\]
Substituting
$\bm w=\bm B_R^{-1}(l\bm e_R-\bm d_R)$ into $r(\bm w)$ gives
\begin{equation}
r(\bm w)
=
1+l E_R-F_R.
\label{eq:r_mu_R}
\end{equation}
Substitution into $q(\bm w)$ gives
\begin{equation}
q(\bm w)
=
l^2E_R-G_R+a_R.
\label{eq:q_mu_R}
\end{equation}
Since $l=q(\bm w)/r(\bm w)$, combining
\eqref{eq:r_mu_R} and \eqref{eq:q_mu_R} yields
\[
l(1+l E_R-F_R)
=
l^2E_R-G_R+a_R.
\]
The $l^2E_R$ terms cancel, giving
\[
l(1-F_R)=a_R-G_R,
\]
which gives the scalar equation for $l$.

We next convert the vector solution into a spike-wise closed form.
From the definitions of $\bm B_R$, $\bm d_R$, and $\bm e_R$,
\[
B_{R,ii}=s_ik_i(1+t_is_i),
\quad
d_{R,i}=s_ik_i(1+t_i),
\quad
e_{R,i}=s_ik_i .
\]
Therefore
\begin{align}
\bar{w}_{i}^*
&=
\frac{
l e_{R,i}-d_{R,i}
}{
B_{R,ii}
} \nonumber\\
&=
\frac{
l-(1+t_i)
}{
1+t_is_i
}.
\label{eq:w_i_mu_proof}
\end{align}
Defining $\chi=l-1$ gives the weight formula in
Theorem~\ref{THRM:DET_OPT}.

It remains to obtain the explicit expression for $\chi$. Since the look direction $\bm a(\theta_1)$ belongs
to the signal-plus-interference spike subspace, and
$k_i = {\bf a}^H(\theta_1){\bf v}_i{\bf v}_i^H{\bf a}(\theta_1)$, we have
\[
\sum_{i=1}^{m}k_i=1.
\]
Using the scalar equation for $l$, $\chi=l-1$ can be written as
\[
\chi
=
\frac{
a_R-1-\bm d_R^T\bm B_R^{-1}\bm d_R
+\bm e_R^T\bm B_R^{-1}\bm d_R
}{
1-\bm e_R^T\bm B_R^{-1}\bm d_R
}.
\]
Substituting the diagonal forms of $\bm B_R$, $\bm d_R$, and
$\bm e_R$ gives
\begin{align}
a_R-1-\bm d_R^T\bm B_R^{-1}\bm d_R
+\bm e_R^T\bm B_R^{-1}\bm d_R
&=
\sum_{j=1}^{m}
\frac{t_jk_j(1-s_j)}
{1+t_js_j},\\
1-\bm e_R^T\bm B_R^{-1}\bm d_R
&=
\sum_{j=1}^{m}
\frac{k_j(1-s_j)}
{1+t_js_j}.
\end{align}
This proves the expression for $\chi$ and therefore the closed-form
robust MVDR-optimal weights in Theorem~\ref{THRM:DET_OPT}.

\section{Proof of Theorem~\ref{THRM:SAMPLE_BASED}}\label{app:proof-sample-based}
The proof has two parts. First, we show that the sample quantities in
Theorem~\ref{THRM:SAMPLE_BASED} consistently estimate the population
quantities used in Theorem~\ref{THRM:DET_OPT}. Second, we apply the
continuous mapping theorem to the closed-form weight formula.

By Couillet's empirical robust spiked random matrix theorem
\cite[Th. 3]{couillet2015robust}, the leave-one-out quantities
defined in Theorem~\ref{THRM:SAMPLE_BASED} consistently replace
the unknown texture distribution $\nu$ in the deterministic robust
spike equations. Consequently, for every detected robust spike,
\begin{equation}
\hat t_i-t_i
\xrightarrow{\rm a.s.}0,
\qquad
\hat s_i-s_i
\xrightarrow{\rm a.s.}0.
\label{eq:proof_sample_s}
\end{equation}

Next, the robust eigenvector overlap relation gives
\[
{\bf a}^H(\theta_1){\bf u}_i{\bf u}_i^H{\bf a}(\theta_1)
\xrightarrow{\rm a.s.}
s_ik_i.
\]
Since 
$\hat s_i\xrightarrow{\rm a.s.}s_i$, we have
\begin{equation}
\hat k_i
=
\frac{1}{\hat s_i}
{\bf a}^H(\theta_1){\bf u}_i{\bf u}_i^H{\bf a}(\theta_1)
\xrightarrow{\rm a.s.}
k_i.
\label{eq:proof_sample_k}
\end{equation}

Construct
\begin{equation}
\hat{\bm B}_R
=
\diag
\left\{
\hat{s}_1 \hat{k}_1(1+\hat{t}_1 \hat{s}_1), \ldots, \hat{s}_m \hat{k}_m(1+\hat{t}_m\hat{s}_m)
\right\},
\label{eq:sample_B_R}
\end{equation}
\begin{equation}
\hat{\bm d}_R
=
\left[
\hat{s}_1 \hat{k}_1(1+\hat{t}_1), \ldots, \hat{s}_m \hat{k}_m(1+\hat{t}_m)
\right],
\label{eq:sample_d_R}
\end{equation}
\begin{equation}
\hat{\bm e}_R
=
\left[
\hat{s}_1 \hat{k}_1, \ldots, \hat{s}_m \hat{k}_m
\right]^T,
\qquad
\hat{a}_R
=
1+\sum_{i=1}^{m}\hat{t}_i\hat{k}_i.
\label{eq:sample_e_a_R}
\end{equation}
Let
\begin{equation}
\hat l
=
\frac{
\hat a_R-\hat{\bm d}_R^T\hat{\bm B}_R^{-1}\hat{\bm d}_R
}{
1-\hat{\bm e}_R^T\hat{\bm B}_R^{-1}\hat{\bm d}_R
},
\label{eq:sample_mu_R}
\end{equation}
and define
\begin{equation}
\hat{\bm w}^\star
=
\hat{\bm B}_R^{-1}
(\hat l\hat{\bm e}_R-\hat{\bm d}_R).
\label{eq:sample_w_R}
\end{equation}

Equations \eqref{eq:proof_sample_s}--\eqref{eq:proof_sample_k}
imply entrywise convergence of the deterministic-equivalent
coefficients:
\begin{equation}
\hat{\bm B}_R\xrightarrow{\rm a.s.}\bm B_R,\qquad
\hat{\bm d}_R\xrightarrow{\rm a.s.}\bm d_R,\qquad
\hat{\bm e}_R\xrightarrow{\rm a.s.}\bm e_R,\qquad
  \hat a_R\xrightarrow{\rm a.s.} a_R
\label{eq:proof_sample_coeff}
\end{equation}
Because $\bm B_R$ is nonsingular and
$1-\bm e_R^T\bm B_R^{-1}\bm d_R\ne0$ by Theorem 2, matrix inversion
and division are continuous in a neighborhood of the limit. Hence
\[
\hat{\bm B}_R^{-1}\xrightarrow{\rm a.s.}\bm B_R^{-1},
\qquad
\hat l\xrightarrow{\rm a.s.}l
\]
Substituting these convergences into \eqref{eq:sample_w_R} gives
\[
\hat{\bm w}^\star
=
\hat{\bm B}_R^{-1}
(\hat l\hat{\bm e}_R-\hat{\bm d}_R)
\xrightarrow{\rm a.s.}
\bm B_R^{-1}(l\bm e_R-\bm d_R)
=
\bm w^\star.
\]
This proves \eqref{w_conv}.

\end{document}